\documentstyle[prl,twocolumn,aps,epsf]{revtex}
\begin{document}

\draft

\tolerance 50000

\twocolumn[\hsize\textwidth\columnwidth\hsize\csname@twocolumnfalse\endcsname

\title{The dimer-RVB State of the Four-Leg Heisenberg Ladder: 
\\ Interference among Resonances}

\author{
M. Roncaglia$^{1}$, G. Sierra$^{1}$ and M.A. Martin-Delgado$^{2}$ 
}
\address{
$^{1}$Instituto de Matem{\'a}ticas y F{\'\i}sica Fundamental, C.S.I.C.,
Madrid, Spain. \\
$^{2}$Departamento de F{\'{\i}}sica 
Te{\'o}rica, Universidad Complutense, Madrid, Spain.
}

\maketitle

\begin{abstract}
\begin{center}
\parbox{14cm}{We study the ground state of the 4-leg spin ladder
using a dimer-RVB ansatz and the Lanczos method. Besides the well known
resonance mechanism between valence bond configurations  
we find novel interference effects among nearby resonances. 
}

\end{center}
\end{abstract}

\pacs{
\hspace{2.5cm}
}

\vskip2pc] \narrowtext

Doped and undoped ladders have focused a lot of attention in recent years
due to the existence of materials with that structure, some of them
are close relatives of the high-$T_c$ oxides 
as the series $Sr_{2n_\ell-2} Cu_{2n_\ell} O_{4n_\ell-2}$  
where $n_\ell$ is the number of legs forming the ladder \cite{T}.  
The undoped Heisenberg spin ladders with $n_\ell$ even 
are known to be spin liquids with 
a spin gap and exponential  decaying correlation 
functions \cite{DR}.
The ground
state (GS) of these low dimensional systems is given by a short-range
RVB ansatz where the topological spin defects are confined \cite{WNS}. 
The RVB  picture is supported by mean field \cite{RGS}, 
DMRG \cite{WNS,WS}, Quantum Montecarlo \cite{QMC} and Lanczos \cite{L}
results  concerning   ladders
with $n_\ell=2,4$ legs and variational 
ansatzs (RVA) for the 2-leg ladder \cite{FM,SM}. 
The purpose of this letter is to apply the RVA method 
to the 4-leg spin ladder with the aim of studying in more detail
the structure of the short-range RVB state. 
In the 2-leg ladder case the basic mechanism 
that  lowers the GS energy is the resonance between two 
nearest neighbour valence bonds \cite{KRS}.
The  simplest short-range RVB  ansatz is given 
by a dimer-RVB state with a single variational 
parameter $u$ which gives the amplitude of the resonance 
\cite{FM,SM}. In the 4-leg ladder case we shall
study  a dimer-RVB ansatz where the resonance may
occur among any possible pair of nearest neigbour parallel 
bonds. The new phenomena
we shall investigate in this letter is the ``interference''  
between couples of resonating bonds. 
We mean by interference  the influence that a pair of
resonating bonds exerts on another pair of near by 
resonating bonds. In the  standard RVB ansatz
of Liang et al. \cite{LDA} the RVB amplitudes have a factorized form
which cannot describe this interference effect.

The Hamiltonian of the 4-leg spin ladder is given by,

\begin{eqnarray}
& H = J  \sum_{a=1}^4 \sum_{n=1}^{N-1}  {\bf S}_a(n) \cdot 
{\bf S}_a(n+1) & \nonumber \\
& + \sum_{n=1}^N \left[ J' ( {\bf S}_1(n) \cdot {\bf S}_2(n) + 
{\bf S}_3(n)\cdot {\bf S}_4(n) ) \right. & \label{1} \\
& \left. + J'' {\bf S}_2(n) \cdot  {\bf S}_3(n) + 
J''' {\bf S}_1(n)\cdot {\bf S}_4(n) \right] & \nonumber 
\end{eqnarray}

\noindent where ${\bf S}_a(n)$ is the spin 1/2 operator at the $a=1, \dots,4$ 
leg and $n=1, \dots, N$ rung. We shall consider the cases of 
periodic or closed boundary conditions (BC) along the rungs, i.e.
$J'=J''= J'''$ and open BC's along the rungs, i.e. 
$J'=J'', J'''=0$. Setting $J''=J'''=0$
we recover two decoupled 2-leg ladder Hamiltonians. 
If $J=0$,  the exact GS of
(\ref{1}) is given by the coherent superposition of the GS of every rung 
which can be written as

\begin{eqnarray}
& |{\rm rung} \rangle = \overline{12}\;\; \overline{34} 
+ u_0 \;  \overline{14} \;\; \overline{32} & \label{2} \\   
& u_0 = \left\{ \begin{array}{ll} 1 & J'=J''=J''' \\
0.366 & J'=J'', J'''=0 \\ 0 & J''=J'''=0 
\end{array} \right. & 
\nonumber 
\end{eqnarray}

\noindent where 
$ \overline{ab}= ( |\uparrow \rangle_a |\downarrow \rangle_b - 
|\downarrow \rangle_a |\uparrow \rangle_b)/\sqrt{2}$
denotes the valence bond state
between the sites  on the legs $a$ and $b$ of the $n^{\rm th}$  rung. 
In figs. 1(a) and 1(b) we depict the valence bond states (\ref{2}).

%%%%%%%%%%%%%%%%%%%%%%%%%%%%%%%%%%%%%%%%%%%%%%%%
\begin{figure}
%\vspace{2.0cm}
\hspace{-0.1cm}
\epsfxsize=7cm \epsffile{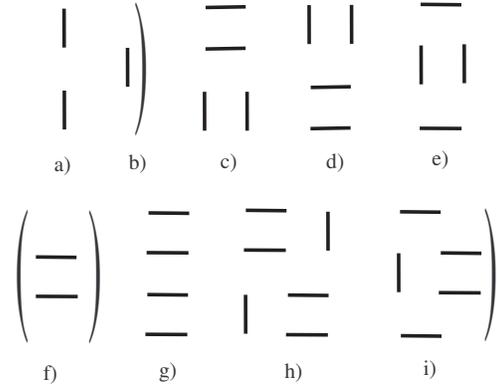}
%\vspace{-3.0cm}
\narrowtext
\caption[]{Graphical representation of the local configurations
that make up the dimer-RVB ansatz for the 4-leg ladder. Every line
connecting two sites $a$ and $b$ corresponds to the valence bond state 
$\overline{ab}$ defined in the text. The site $a$ belongs  
to the even sublattice while the site $b$ belongs to the odd one.  
}
\label{fig1}
\end{figure}
%%%%%%%%%%%%%%%%%%%%%%%%%%%%%%%%%%%%%%%%%%%%%%%%%

Switching on the intraleg coupling $J$ 
any  pair of rung-bonds will start to resonate
with a pair of leg-bonds as in figs 1(c,d,e,f). 
There are 4 types of ``elementary'' resonances
involving two consecutive rungs $n$ and $n+1$ 
and two legs $i$ and $j$, which we denote as   
$(12), (34), (14)$ and $(23)$. We associate 
an amplitude  $u_{ij}$ to every of these resonances. 
There is also a state with 4 leg-bonds 
on two consecutive rungs, which we denote
as $(1234)$, and give it  an amplitude $u_{1234}$ 
(see fig. 1(g)).  Finally, 
we  may have two resonances $(ij)$ and $(kl)$ sharing a common rung
as in figs. 1(h,i), which we  
denote as  $(12,34)$ and $(14,23)$, and give them amplitudes 
$u_{12,34}=u_{34,12}$ and $u_{14,23}=u_{23,14}$ 
respectively. Let us suppose that a pair of 
resonating bonds is not influenced by its environement.
This would imply the following factorization 
$u_{ij,kl}= u_{ij} u_{kl}$, which as we shall see 
below never happens. 
Figs. 1 display  all the local configurations 
that should be combined in all possible manners to produce a
dimer-RVB ansatz. This seems to be a formidable problem 
if we try to solve
it with standard combinatorial methods. However, as in the 2-leg ladder
case \cite{SM},  the dimer-RVB state of the 4-leg ladder can be 
generated by the set of recurrence relations  (RR) 
given in fig.2.

%%%%%%%%%%%%%%%%%%%%%%%%%%%%%%%%%%%%%%%%%%%%%%%%
\begin{figure}
%\vspace{2.0cm}
\hspace{-0.1cm}
\epsfxsize=7cm \epsffile{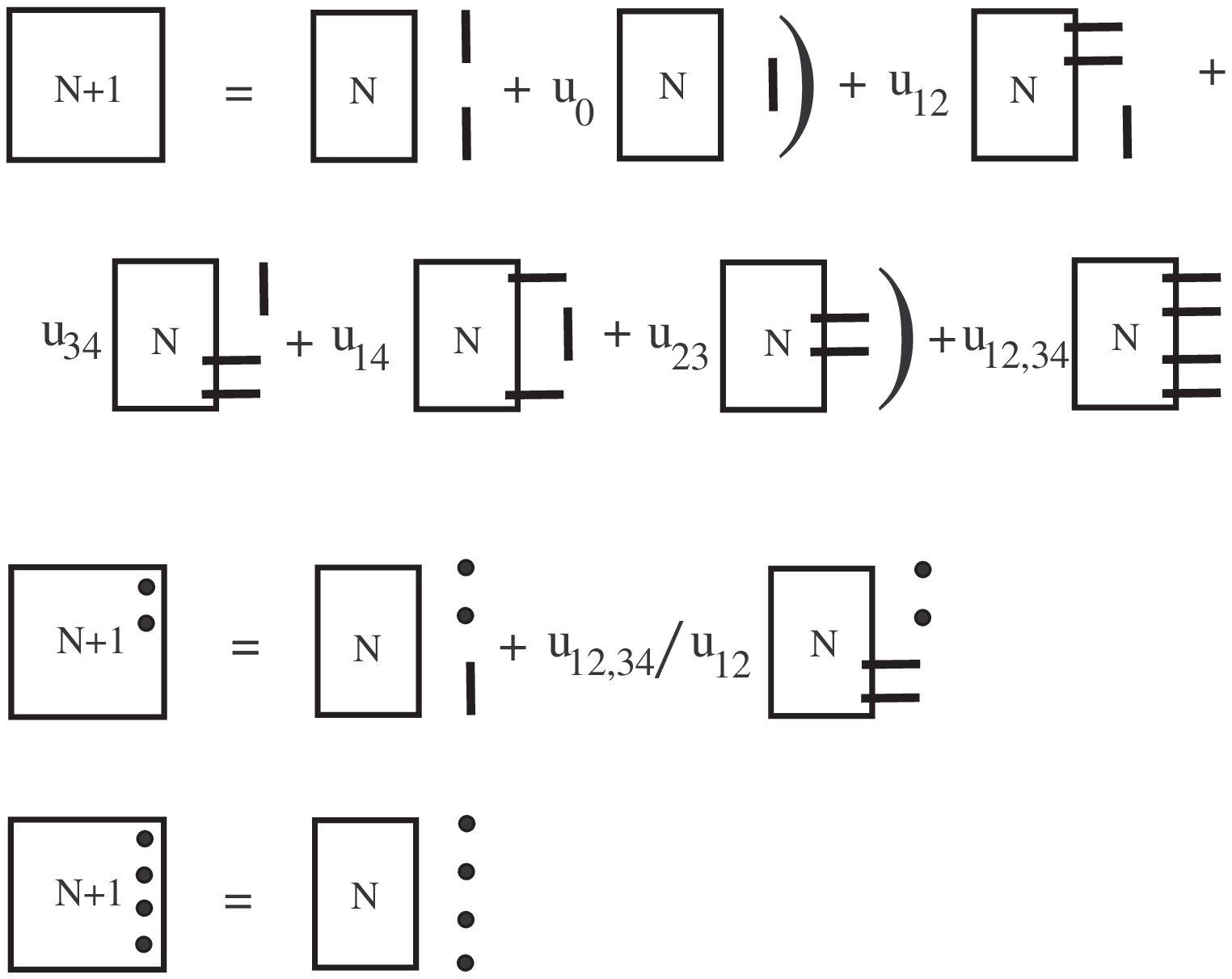}
%\vspace{-3.0cm}
\narrowtext
\caption[]{Recurrence relations that generate iteratively
the dimer-RVB state of the 4-leg ladder. The empty box
represents the singlet state  $|N\rangle$ 
of the ladder with $N$ rungs.
A box  with two dots on the legs $i$ and $j$ 
represents the state 
$|N, \sigma_i, \sigma_j \rangle$, where 
$\sigma_i$ and $\sigma_j$ are  free spins which form valence
bonds with nearest neighbour spins located in the same legs. 
We give explicitely the RR of the state 
$|N, \sigma_1, \sigma_2 \rangle$. The RR's of the other
two-dotted states are similar. The last RR is that
of the 4-dotted state
$|N, \sigma_1, \sigma_2, \sigma_3, \sigma_4 \rangle$.}
\label{fig2}
\end{figure}
%%%%%%%%%%%%%%%%%%%%%%%%%%%%%%%%%%%%%%%%%%%%%%%%%

\noindent  
Fig.3 shows a state generated by these RR's. One can characterize 
a dimer state with $N$ rungs by the collection  of legs 
that one  cuts between two consecutive rungs.
If no legs are cut we write $(0)$, cutting the  legs $i$ and $j$
we write $(ij)$ and cutting  four legs we write $(1234)$. 
With  these notations 
the state of  fig.3. reads  $(12)(0)(34)(12)(0)(23)$
and has an amplitude $u_{12}  u_{12,34} u_{12} u_{23}$.

It is important to realize that not all the states of the form 
$A_{12} A_{23} A_{34} \dots A_{N-1,N}$ (where $A_{n,n+1}$ 
denotes the set
of legs cut between  the rungs $n$ and $n+1$)  are allowed. 
For example, after the configuration $(12)$ one can only have
either $(0)$ or $(34)$, or  after $(1234)$ only $(0)$ may follows. 
These selection rules are  summarized in the graph
of  fig. 4. The vertices  of the graph  denote
the configurations $A= (0), (12), (34), (14), (23), (1234)$ 
while a link between the vertices  $A$ and $A'$ indicates
that these two configurations may  appear consecutively 
in an allowed dimer state. The amplitudes of the dimer
states are associated to the links of the graph.

%%%%%%%%%%%%%%%%%%%%%%%%%%%%%%%%%%%%%%%%%%%%%%%%
\begin{figure}
%\vspace{2.0cm}
\hspace{1cm}
\epsfxsize=5cm \epsffile{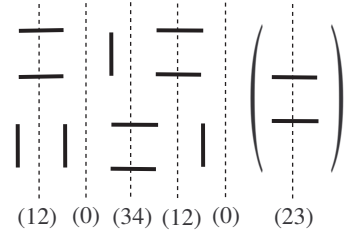}
%\vspace{-3.0cm}
\narrowtext
\caption[]{A dimer state constructed with the RR's given
in fig.2. The dotted lines represents the cuts  described
in the text. 
}
\label{fig3}
\end{figure}
%%%%%%%%%%%%%%%%%%%%%%%%%%%%%%%%%%%%%%%%%%%%%%%%%

%%%%%%%%%%%%%%%%%%%%%%%%%%%%%%%%%%%%%%%%%%%%%%%%
\begin{figure}
\vspace{-1.0cm}
\hspace{2cm}
%\begin{center}
\epsfxsize=3.5cm \epsffile{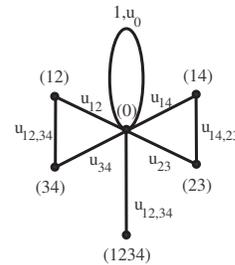}
%\vspace{-3.0cm}
\narrowtext
\caption[]{Graph that encodes the  dimer configurations
of the 4-leg ladder. The vertices are labelled by the legs
cut between two consecutive rungs. A link between two
vertices represents cuts that share a common rung.   
Every link is associated with a variational parameter 
of the RR's. The link connecting  $(0)$ to itself
means  that the  middle rung between the two cuts 
is a singlet which may be either 
$\overline{12} \;\; \overline{34}$, with amplitude
1 or $\overline{14} \;\; \overline{23}$, with amplitude
$u_0$.}
\label{fig4}
%\end{center}
\end{figure}
%%%%%%%%%%%%%%%%%%%%%%%%%%%%%%%%%%%%%%%%%%%%%%%%%

The RR's of fig.2 generate 
all the dimer states of a 4-leg ladder
with periodic BC's along the rungs, and their number
grows exponentially with the number of legs \cite{dimer}. 
For the open BC's we should restrict
ourselves to dimer states with no bonds of length greater
than one. However  the strong coupling 
limit $J/J' <<1$ forces to include the valence 
bond $\overline{14}$ as in eq.(\ref{2}).  So the distinction
between closed and open ladders will only appear  in the
variational parameters. 
The existence of RR's to generate the GS ansatz implies that the norm and
expectation value of the Hamiltonian (\ref{1}) also satisfies RR's, which can
be iterated to give the energy 
of the ansatz $\langle H  \rangle_N$  for any number
of rungs $N$. The set of variational 
parameters ${ u_X}$ is obtained by minimization of
$\langle H \rangle_N $. 
This method is similar to 
the matrix product ansatz of references \cite{MP}
 but differs in that the states 
kept are non-orthogonal as corresponds to a RVB ansatz.
Let us next present our results.

%%%%%%%%%%%%%%%%%%%%%%%%%%%%%%%%%%%%%%%%%%%%%%%%
\begin{figure}
\vspace{2.0cm}
\hspace{1.cm}
\epsfxsize=5.5cm \epsffile{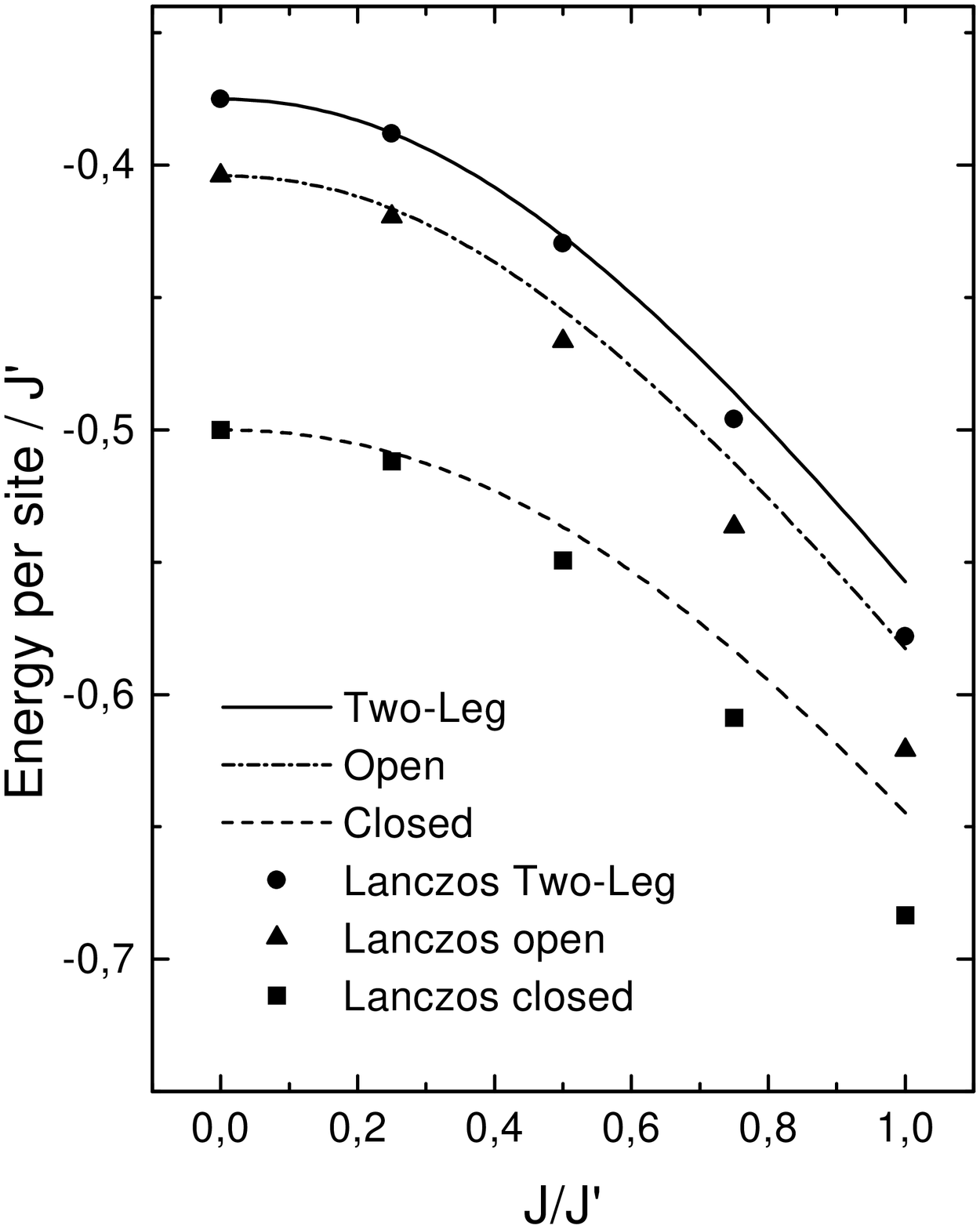}
%\vspace{-3.0cm}
\narrowtext
\caption[]{GS energy per site in units of $J'$ 
of the 4-leg dimer-RVB state with closed and open
BC's and the 2-leg ladder in the range $0< J/J' <1$. 
We also plot the exact GS energies obtained by 
extrapolating Lanzcos results to the thermodynamic limit with ladders
of sizes $4 \times n ( n=4,5,6,7)$. 
}
\label{fig5}
\end{figure}
%%%%%%%%%%%%%%%%%%%%%%%%%%%%%%%%%%%%%%%%%%%%%%%%%

In fig. 5 we plot the GS energy 
per site obtained with our variational ansatz  and  
the Lanczos method in the range of   
couplings $0 < J/J'<1$. We also include  
for comparison the GS energy per site of the 
2-leg ladder. The GS energies are 
very close to the exact result in the strong
coupling region $0 <J/J'< 0.3 $, but they 
get worse for larger couplings. This is
natural since configurations with longer
bonds are expected to become more
important in the weak interleg coupling regime.
The closed-rung ladder has a much lesser 
GS energy per site than the open one. This is mainly 
due to the resonance (\ref{2}) between the
two bonds along  the rungs. 
The GS curves for open and closed ladders  in fig. 5 can be fitted
with the formula,

\begin{eqnarray}
& E_0(N)/(4N J') = - e_0 - e_1 (J/J')^2 -e_2 (J/J')^4 & \label{3} \\ 
&  (e_0,e_1,e_2) = \left\{
\begin{array}{llll} 0.5, & 0.15, & -0.005, & {\rm closed}  
\\ 0.404, &0.23,& -0.05 & {\rm open} \end{array}
\right. & \nonumber 
\end{eqnarray}

\noindent where $e_0$ is the energy per site of a single rung. 
Eq.(\ref{3}) agrees with perturbation theory up 
to second  order.

Let us consider now the behaviour of the variational parameters.  
In the closed-rung 
case the choice of couplings $J'=J''=J'''$ implies the existence of
a rotational symmetry among  the legs which leaves  
only 4 independent variational parameters given by,

\begin{equation}
u_0,\;  u_{12}=u_{ij},\; v_{12,34} = 
\frac{u_{ij,kl}}{u_{ij} u_{kl}}, \; u_{1234}
\label{4}
\end{equation}

In fig. 6 we plot these parameters  in the 
domain $0 < J/J'<1$. Let us comment on these results.

\begin{itemize}

\item $u_0$ takes the constant value 1, which coincides with the
exact $J=0$ result (\ref{2}). This implies the  absence of interference 
between rung and leg resonance.

%%%%%%%%%%%%%%%%%%%%%%%%%%%%%%%%%%%%%%%%%%%%%%%%
\begin{figure}
%\vspace{2.0cm}
\hspace{1.cm}
\epsfxsize=5.cm \epsffile{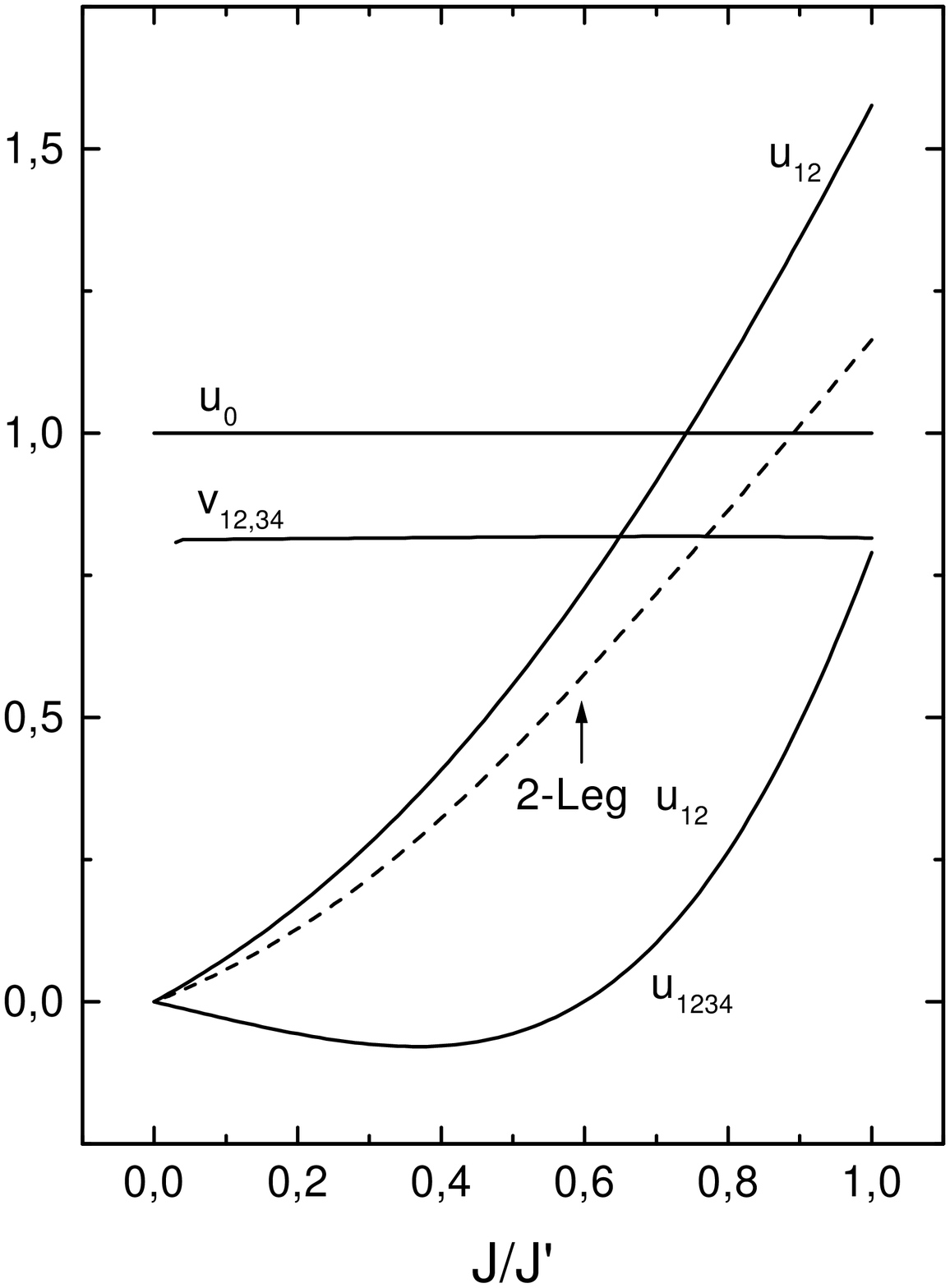}
%\vspace{-3.0cm}
\narrowtext
\caption[]{Variational parameters (\ref{4}) 
of the closed BC dimer-RVB in the range 
$0 < J/J' <1$.  We include for comparison the 
value of the variational parameter $u$ of the
2-leg ladder \cite{SM}. 
}
\label{fig6}
\end{figure}
%%%%%%%%%%%%%%%%%%%%%%%%%%%%%%%%%%%%%%%%%%%%%%%%%

\item $u_{12}$ is greater 
than its 2-leg analog $u$ \cite{SM}. For the isotropic
case one gets $u_{12}= 1.58$ while $u=1.18$ \cite{SM}. 
Simple resonance is enhanced 
in the 4-leg ladder. 

\item $v_{12,34}$ is almost constant and less than one 
indicating destructive interference
between resonances shearing a common rung as in fig.1(c).

\item $u_{1234}$ displays an unexpected  behaviour
since it first becomes negative for small values of $J/J'$,  reaches a minimum
and starts to grow becoming positive for $J/J' > 0.6$. 
This peculiar behaviour of $u_{1234}$ is  
a sign of destructive interference between resonances
sharing two rungs. 

\end{itemize}

In the case of open ladders, $J''=J', J'''=0$, one is left with 7 independent
variational parameters given by,

\begin{eqnarray}
& u_0,\; u_{12}= u_{34}, \;  u_{14}, \; u_{23}, \;
\;v_{ij,kl} = \frac{u_{ij,kl}}{u_{ij} u_{kl}}, 
\; u_{1234} & \label{5}
\end{eqnarray}

In fig. 7 we plot the values of these parameters in the range
$0 < J/J' <1$.  Some features that we encounter in 
fig.7 have already appeared in the periodic case.

\begin{itemize}

\item $u_0$ stays almost constant with a value close to the
exact  
$J=0$ result (\ref{2}).

\item $u_{12}$ and $u_{14}$ are quite similar,  but 
$u_{23}$ is much smaller. So bonds do not like
to resonate in the middle of the ladder. This
is due to lost of energy induced by the 
existence of the  long bond $\overline{14}$.

\item $v_{12,34}$ is lower than 1, as in the periodic
case,  but $v_{14,23}$ is much greater than 1, which
is again due to the smallness of $u_{23}$. For graphical
purposes we plot in fig.7 the inverse of $v_{14,23}$.

%%%%%%%%%%%%%%%%%%%%%%%%%%%%%%%%%%%%%%%%%%%%%%%%
\begin{figure}
%\vspace{2.0cm}
\hspace{1.cm}
\epsfxsize=5cm \epsffile{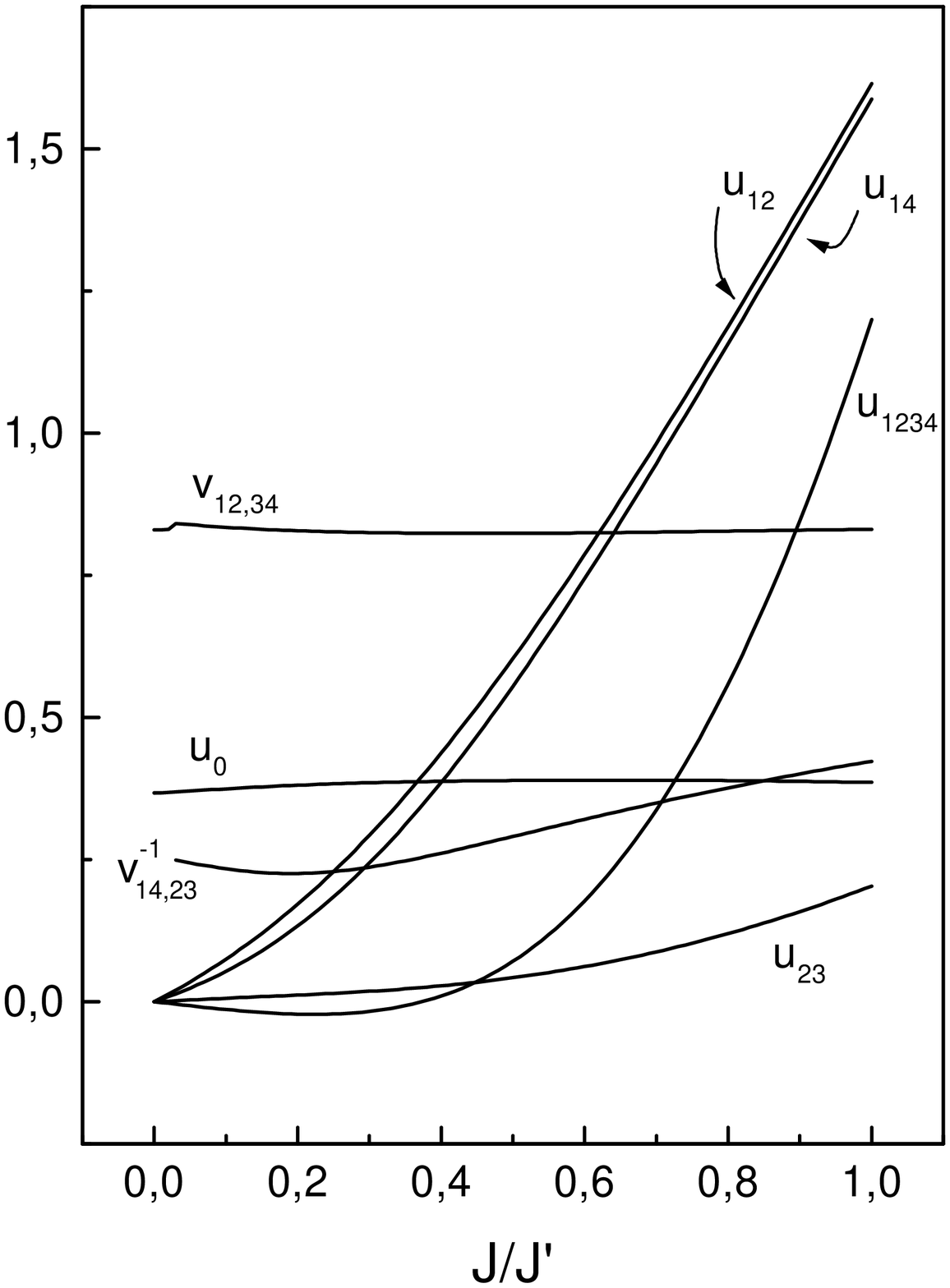}
%\vspace{-3.0cm}
\narrowtext
\caption[]{Variational parameters (\ref{5}) 
of the open BC dimer-RVB in the range 
$0 < J/J' <1$.
}
\label{fig7}
\end{figure}
%%%%%%%%%%%%%%%%%%%%%%%%%%%%%%%%%%%%%%%%%%%%%%%%%

\item $u_{1234}$ is also  suppressed but in a smaller amount
than in the periodic ladder.

\end{itemize}

We have also computed the spin correlation length $\xi$
from the exponential decaying behaviour of the spin-spin correlator.
For the isotropic case we get $\xi=0.81$ for the closed ladder
and $\xi=0.92$ for the open one. These results show that the
rung configurations are more important for the closed ladder
than for the open one, which is in agreement with the values
taken by the variational parameters. The DMRG 
method yields $\xi=5 \sim 10$ \cite{WNS} for open  ladders,
while QMC method yields $\xi=7.1$ (closed) and 
$\xi=10.3$ (open) \cite{QMC}.
As expected the dimer-RVB ansatz gives a much shorter correlation
length but it  reproduces the fact that $\xi_{\rm closed} <
\xi_{\rm open}$.

We have also studied
the case when the Hamiltonian (\ref{1}) becomes that of two
decoupled 2-leg ladders, i.e. $J''=J'''=0$. Curiously enough
our ansatz yields a GS with bonds connecting the two ladders.
The GS energy so obtained is a bit lower than the one of two
uncoupled 2-leg ladders and the correlation length $\xi=0.97$
is larger than in the uncoupled case $\xi=0.79$ \cite{SM}.

In summary we have shown in this letter  that the dimer-RVB ansatz
gives a correct qualitative picture of the short-range RVB state
of the 4-leg ladder. We have found 
interesting  interference effects between  resonating
valence bond configurations which should probably carry over 
more realistic ansatzs which must  include longer
valence bonds. 
The next step is to generalize our   methods 
 to the  doped 4-leg ladders
where one can study the phenomena of  phase-separation 
and stripe formation \cite{WS,D}. Previous applications
of the RVA method to
the 2-leg $t-J$ ladder
\cite{holes}, the necklace $t-J$ ladder \cite{necklace}
and the 2-leg Hubbard model \cite{kim}, suggest that this
goal is worth pursuing.

{\bf Acknowledgments} 
We would like to thank J. Dukelsky for conversations
and the  Centro de Supercomputacion Complutense for the 
allocation of CPU time in the SG-Origin 2000 Paralel Computer.
This work was supported by the DGES spanish grants
PB97-1190 (G.S. and M.A.M.-D.).

%%%%%%%%%%%%%%%%%%%%%%%%%%%%%%%%%%%%%%%

%%%%%%%%%%%%%%%%%%%%%%%%%%%%%%%%%%%%%%%

\end{document}